\documentclass[12pt,preprint]{aastex}
\usepackage[dvips]{color}
\usepackage{graphicx}
\usepackage{ulem}

\def\msol{\hbox{\kern 0.20em $M_\odot$}}
\def\lsol{\hbox{\kern 0.20em $L_\odot$}}
\def\rsol{\hbox{\kern 0.20em $R_\odot$}}
\def\sr{\hbox{\kern 0.20em sr}}
\def\srmu{\hbox{\kern 0.20em sr$^{-1}$}}
 
\def\g{\hbox{\kern 0.20em g}}
\def\gmu{\hbox{\kern 0.20em g$^{-1}$}}
\def\kg{\hbox{\kern 0.20em kg}}
\def\pc{\hbox{\kern 0.20em pc}}
 
\def\mum{\hbox{\kern 0.20em $\mu$m}}
\def\mumd{\hbox{\kern 0.20em $\mu$m$^{-2}$}}
\def\cm{\hbox{\kern 0.20em cm}}
\def\m{\hbox{\kern 0.20em m}}
\def\km{\hbox{\kern 0.20em km}}
\def\nm{\hbox{\kern 0.20em nm}}
 
\def\s{\hbox{\kern 0.20em s}}
\def\h{\hbox{\kern 0.20em h}}
\def\sec{\hbox{\kern 0.20em sec}}
\def\min{\hbox {\kern 0.20em min}}
\def\smu{\hbox{\kern 0.20em s$^{-1}$}}
\def\smd{\hbox{\kern 0.20em s$^{-2}$}}
\def\an{\hbox{\kern 0.20em an}}
\def\anmu{\hbox{\kern 0.20em an$^{-1}$}}
\def\deg{\hbox{\kern 0.20em $^{\rm o}$}}
\def\yr{\hbox{\kern 0.20em yr}}
\def\yrmu{\hbox{\kern 0.20em yr$^{-1}$}}
\def\Myr{\hbox{\kern 0.20em Myr}}
\def\Mymu{\hbox{\kern 0.20em Myr$^{-1}$}}
\def\K{\hbox{\kern 0.20em K}}
\def\pcmu{\hbox{\kern 0.20em pc$^{-1}$}}
\def\pcmd{\hbox{\kern 0.20em pc$^{-2}$}}
\def\pcmt{\hbox{\kern 0.20em pc$^{-3}$}}
\def\kms{\hbox{\kern 0.20em km\kern 0.20em s$^{-1}$}}
\def\kmpd{\hbox{\kern 0.20em km$^{2}$}}
\def\kpc{\hbox{\kern 0.20em kpc}}
\def\cms{\hbox{\kern 0.20em cm\kern 0.20em s$^{-1}$}}
\def\erg{\hbox{\kern 0.20em erg}}
\def\ergs{\hbox{\kern 0.20em erg}}
\def\cmpd{\hbox{\kern 0.20em cm$^2$}}
\def\cmmd{\hbox{\kern 0.20em cm$^{-2}$}}
\def\cmms{\hbox{\kern 0.20em cm$^{-6}$}}
\def\cmpt{\hbox{\kern 0.20em cm$^3$}}
\def\cmmt{\hbox{\kern 0.20em cm$^{-3}$}}
\def\mpd{\hbox{\kern 0.20em m$^2$}}
\def\mmd{\hbox{\kern 0.20em m$^{-2}$}}
\def\mpt{\hbox{\kern 0.20em m$^3$}}
\def\mmt{\hbox{\kern 0.20em m$^{-3}$}}
\def\mujy{\hbox{\kern 0.20em $\mu$Jy}}
\def\mjy{\hbox{\kern 0.20em mJy}}
\def\Mj{\hbox{\kern 0.20em MJy}}
\def\jy{\hbox{\kern 0.20em Jy}}
\def\ghz{\hbox{\kern 0.20em GHz}}
\def\srmd{\hbox{\kern 0.20em sr$^{-1}$}}
\def \kms{km~$\rm{s}^{-1}$}

\def \mum{$\mu$m}

\def\G{\hbox{\kern 0.20em G}}

\def\h13cop{\hbox{H$^{13}$CO$^{+}$}}

\def\S+{\hbox{S{\small II}}}

\newcommand{\myemail}{galliano@astro.umd.edu}

\newcommand{\seppar}{\vspace*{10pt}}

\newcommand{\sms}[1]{{\mbox{{\scriptsize #1}}}}

\newcommand{\reffig}[1]{Fig.~\ref{#1}}
\newcommand{\reffigs}[1]{Figs.~\ref{#1}}
\newcommand{\refsec}[1]{Sect.~\ref{#1}}

\newcounter{textlistctr}

\newcommand{\squishlist}{
   \begin{list}{$\bullet$}
    { \setlength{\itemsep}{0pt}      \setlength{\parsep}{3pt}
      \setlength{\topsep}{3pt}       \setlength{\partopsep}{0pt}
      \setlength{\leftmargin}{1.5em} \setlength{\labelwidth}{1em}
      \setlength{\labelsep}{0.5em} } }

\newcommand{\squishlisttwo}{
   \begin{list}{$\bullet$}
    { \setlength{\itemsep}{0pt}    \setlength{\parsep}{0pt}
      \setlength{\topsep}{0pt}     \setlength{\partopsep}{0pt}
      \setlength{\leftmargin}{2em} \setlength{\labelwidth}{1.5em}
      \setlength{\labelsep}{0.5em} } }

\newcommand{\squishend}{
    \end{list}  }

\newcounter{obsrefctr}
\newcommand{\IC}[1]{IC$\;$#1}

\newcommand{\M}[1]{M$\;$#1}

\newcommand{\ngc}[1]{NGC$\;$#1}

\newcommand{\um}[1]{UM$\;$#1}

\newcommand{\orb}{Orion bar}

\newcommand{\xxxdor}{30$\;$Doradus}

\newcommand{\mic}{\;\mu {\rm m}}

\newcommand{\hii}{H$\,${\sc ii}}

\newcommand{\neiii}{Ne$\,${\sc iii}}
\newcommand{\neii}{Ne$\,${\sc ii}}

\newcommand{\neiiiline}{[\neiii]$_{15.56\mu m}$}
\newcommand{\neiiline}{[\neii]$_{12.81\mu m}$}

\newcommand{\iso}{{\it ISO}}

\newcommand{\spitz}{{\it Spitzer}}

\newcommand{\sofia}{{\it Sofia}}
\newcommand{\jwst}{{\it JWST}}

\newcommand{\irs}{{\it Spitzer}/IRS}
\newcommand{\isocam}{{\it ISO}/CAM}
\newcommand{\isosws}{{\it ISO}/SWS}

\newcommand{\ipah}[1]{{\rm I}_{#1}}
\newcommand{\iPah}{{\rm I}_\sms{PAH}}
\newcommand{\icont}{{\rm I}_\sms{cont}}

\slugcomment{The Evolving ISM in the Milky Way \& Nearby Galaxies}

\shorttitle{PAH Variations in Galaxies}
\shortauthors{F. {\c Galliano}}

\begin{document}

\title{A MULTISCALE STUDY OF POLYCYCLIC AROMATIC HYDROCARBON PROPERTIES 
       IN GALAXIES}

\author{Fr\'ed\'eric {\sc Galliano}\altaffilmark{1}}

\altaffiltext{1}{Department of Astronomy, University of Maryland, 
                 College Park, MD 20742 ({\tt\myemail})}

\begin{abstract}
In the present contribution, I summarize a systematic study of
\iso\ and \spitz\ mid-IR spectra of Galactic regions and star forming galaxies. 
This study quantifies the relative variations of the main aromatic features 
inside spatially resolved objects as well as among the integrated spectra of 50 
objects.
Our analysis implies that the properties of the PAHs are remarkably universal 
throughout our sample and at different spatial scales.
In addition, the relative variations of the band ratios, as 
large as one order of magnitude, are mainly controled by the fraction of ionized 
PAHs. 
In particular, I show that we can rule out both the modification of the PAH 
size distribution and the mid-IR extinction, as an explanation of these 
variations.
High values of the $\ipah{6.2}/\ipah{11.3}$ ratio are found to be associated with 
the far-UV illuminated surface of PDRs, at the scale of an interstellar cloud,
and associated with star formation activity, at the scale of a galaxy.
Using a few well-studied Galactic regions, we provide an empirical relation 
between the $\ipah{6.2}/\ipah{11.3}$ ratio and the ionization/recombination 
ratio $G_0/n_e\sqrt{T_\sms{gas}}$. 
Finally, I show that these trends are consistent with the detailed modeling
of the PAH emission within photodissociation regions, taking into account the
radiative transfer, the stochastic heating and the charge exchange between
gas and dust.
\end{abstract}


\lefthead{F. {\sc Galliano}}

\righthead{PAH Variations in Galaxies}


\section{INTRODUCTION}

Polycyclic Aromatic Hydrocarbons (PAHs) are large molecules containing 
$\sim$10 to 1000 carbon atoms.
They are commonly believed to be the carriers of the ubiquitous broad 
mid-infrared features, centered around 3.3, 6.2, 7.7, 8.6 and 11.3$\mic$ 
\citep[{e.g.}][]{allamandola99}.
Due to their small size, these molecules are predomantly excited by single 
UV-photon events.
At solar metallicity, $\sim15\%$ of the total infrared (IR) luminosity is 
radiated through the aromatic features \citep[{e.g.}][]{zubko04}.
On the contrary, PAHs are underabundant in low-metalicity environments 
\citep[][for a complete study]{galliano08a}.

It has long been surmised that the mid-infrared emission features provide a 
clear signature of the interaction of far-UV photons with cloud surfaces and 
hence a probe of the importance of massive star formation in a region 
\citep{genzel98,peeters04}.
The present work extends this by studying methodically the mid-infrared spectra
of a wide variety of sources at different spatial scales, 
and investigating the physical origin of the PAH band ratio variations.


\begin{figure}[htbp]
  \centering
  \includegraphics[width=\textwidth]{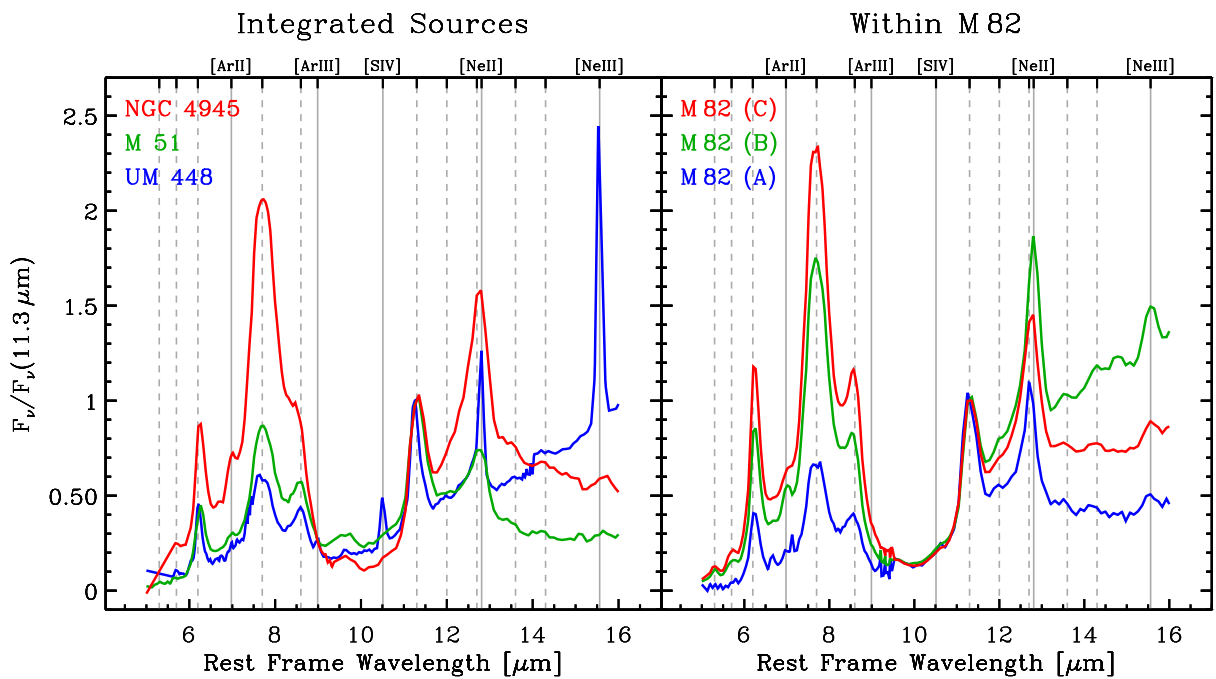}
  \caption{{\footnotesize\sl 
            Qualitative demonstration of the aromatic feature variations in
            galaxies \citep{galliano08b}.
            Both panels show mid-infrared spectra normalised by the 
            monochromatic flux in the 11.3$\mic$ feature.
            \uline{Left panel:} integrated spectra of three galaxies; a blue
            compact dwarf, \um{448}; a spiral \M{51}; a ULIRG \ngc{4945}.
            \uline{Right panel:} spatially resolved spectra in three different
            regions in the starburst galaxy \M{82}; (A) is in the halo; 
            (B) is at an end of the disc; (C) is in the starburst region.}}
  \label{fig:specexpl}
\end{figure}

\section{SYSTEMATIC ANALYSIS OF THE OBSERVED AROMATIC FEATURES}

  \subsection{Sample and Mid-Infrared Spectral Decomposition}

We have constructed a large sample of mid-infrared spectra of Galactic regions
(\hii\ regions, photodissociation regions, planetary nebulae), Magellanic 
\hii\ regions, and nearby dwarf, spiral and starburst galaxies.
These sources were observed with one of the three following instruments:
\isocam, \isosws, \irs.
Several of these objects have been spectrally mapped (\IC{342}, 
\M{17}, \M{51}, \M{82}, \M{83}, \xxxdor, and the \orb).
Our sources are the merging of the samples presented by \citet{madden06} and
\citet{galliano08a}.
They present a wide range of properties (metallicity, star formation rates), 
as well as of spatial scales (resolution of $\sim0.1$~pc in \M{17} and the \orb, 
and of $\sim 0.1$~kpc in external galaxies).
\reffig{fig:specexpl} illustrates the diversity of these properties.

\begin{figure}[htbp]
  \centering
  \includegraphics[width=\textwidth]{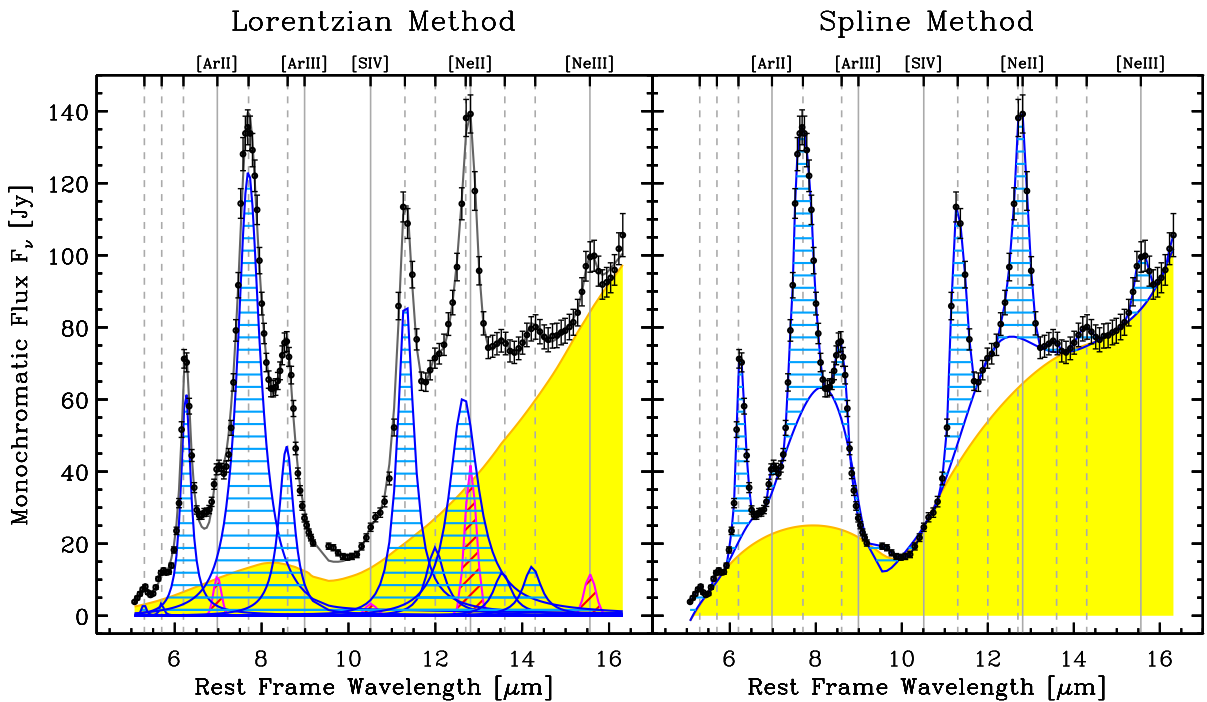}
  \caption{{\footnotesize\sl 
            Demonstration of the two methods used to decompose all our mid-IR 
            spectra, on the spectrum of \M{82} \citep{galliano08b}.
            \uline{Left panel:}
            the continuum (yellow filled) is the sum of two modified
            black-bodies;
            the aromatic features (blue shaded) are Lorentz profiles;
            the ionic lines (pink shaded) are gaussian functions.
            \uline{Right panel:}
            the continuum (yellow filled) is a spline function;
            the plateaus under the bands (empty areas) are also spline 
            functions;
            the aromatic bands and ionic lines (blue shaded) are the residuals.
            }}
  \label{fig:method}
\end{figure}

Quantifying the spectral variations demonstrated in \reffig{fig:specexpl}
is not straightforward.
Indeed, a significant fraction of the power radiated through the aromatic 
features comes from their wings.
Moreover the accurate profile of the bands is uncertain, and has been modeled
differently by \citet{boulanger98}, \citet{li01} and \citet{vermeij02}.
In order to test the robustness of our results, we have systematically analysed 
our spectra using two different methods (\reffig{fig:method}).
These methods are described in detail by \citet{galliano08b}.
It is important to note that the {\it Lorentzian method} (left panel of 
\reffig{fig:method}) fits Lorentz profiles to the aromatic features, taking
into account the flux in their wings.
This method is feature-biased, since part of the small grain continuum can
artificially be accounted by the wings of the bands.
On the contrary, the {\it Spline method} (right panel of \reffig{fig:method}),
which integrates only the tip of the aromatic bands, is continuum-biased.

Although the numerical values of the intensity of a given feature, measured
with the two methods are different, our study shows that the order between 
several intensities is identical with the two methods.
It indicates that the respective biases of the methods do not affect the general
trends between band intensities.
Therefore, in the following of this review, I will present only the results 
obtained with the first method and refer the reader to \citet{galliano08b} 
for the complete results.

\begin{figure}[htbp]
  \centering
  \includegraphics[width=\textwidth]{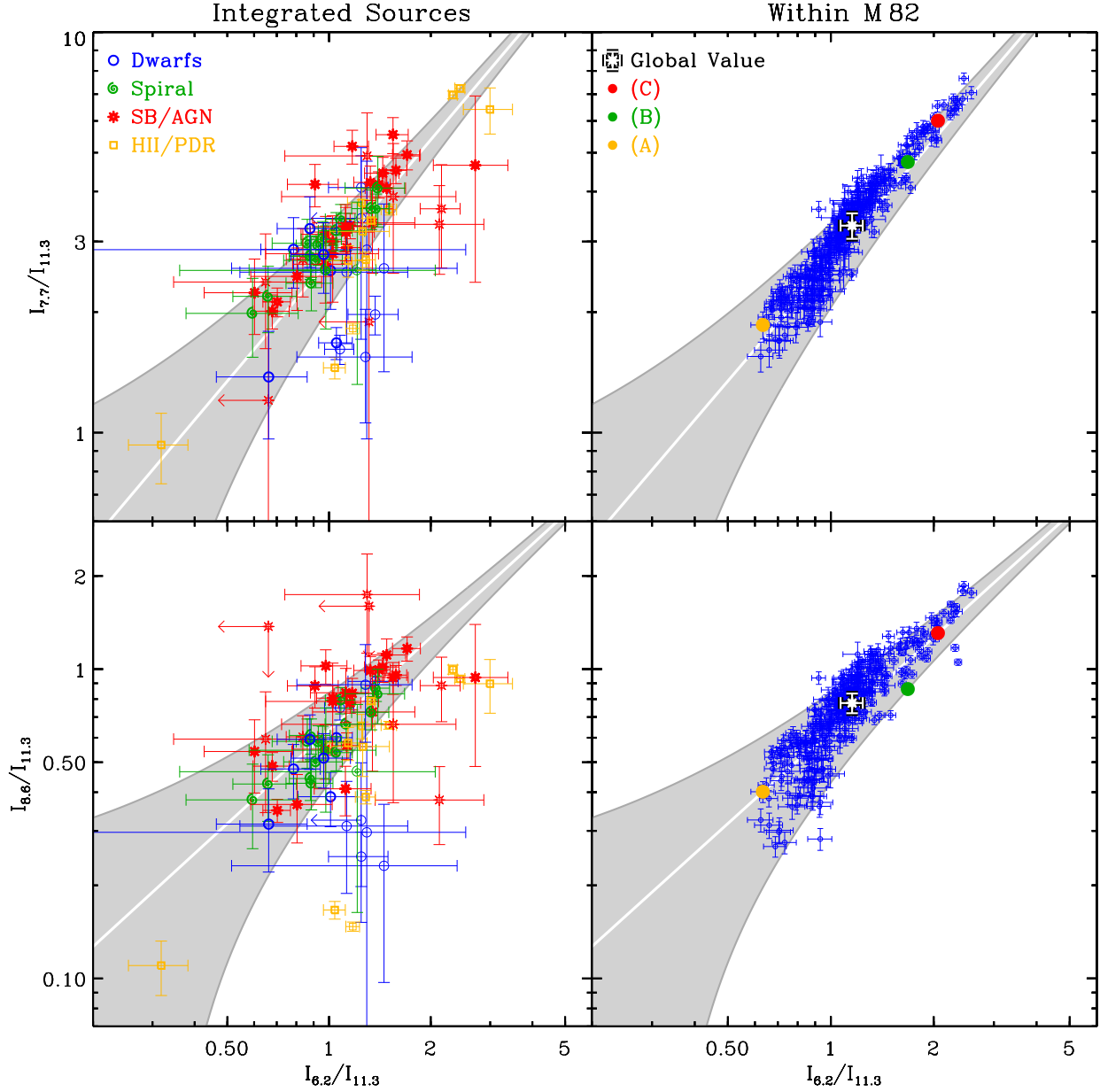}
  \caption{{\footnotesize\sl 
            Correlations between select aromatic feature ratios 
            \citep{galliano08b}.
            $\ipah{\lambda}$ refers to the intensity of the band centered
            around $\lambda\mic$.
            \uline{Left panels:} the symbols correspond to the band ratios 
            derived from spectra integrated over an entire source;
            the faint symbols are fits that we consider uncertain, due either
            to a low signal-to-noise ratio or to a weak feature-to-continuum
            ratio.
            \uline{Right panels:} the symbols correspond to the band ratios
            derived from spectra of individual pixels within \M{82};
            the three regions (A, B, C) are the ones shown in 
            \reffig{fig:specexpl}.
            In all panels, the grey stripe is the linear correlation to the
            observations $\pm1\times\sigma$.
            }}
  \label{fig:correl}
\end{figure}

  \subsection{Correlations Between Aromatic Feature Intensity Ratios}

\reffig{fig:correl} shows the correlations between several aromatic feature
intensity ratios derived from integrated spectra of galaxies and Galactic 
regions (left panels), as well as from spatially resolved spectra within 
\M{82} (right panels).
These figures clearly show that all types of sources and all types of apertures
follow the same universal trends.
The 6.2, 7.7 and 8.6$\mic$ features appear to be tied together, while the ratios
between these features and the 11.3$\mic$ band vary by one order of magnitude.
The relations involving the 8.6$\mic$ feature exhibit a larger dispersion, as
this band is the weakest of the four considered here, its intensity is 
significantly affected by the silicate extinction feature, and it is blended 
with the powerful 7.7$\mic$ feature.
A preliminary version of these correlations was presented by \citet{galliano04}.

\begin{figure}[htbp]
  \centerline{
  \begin{tabular}{cc}
    \includegraphics[width=0.5\textwidth]{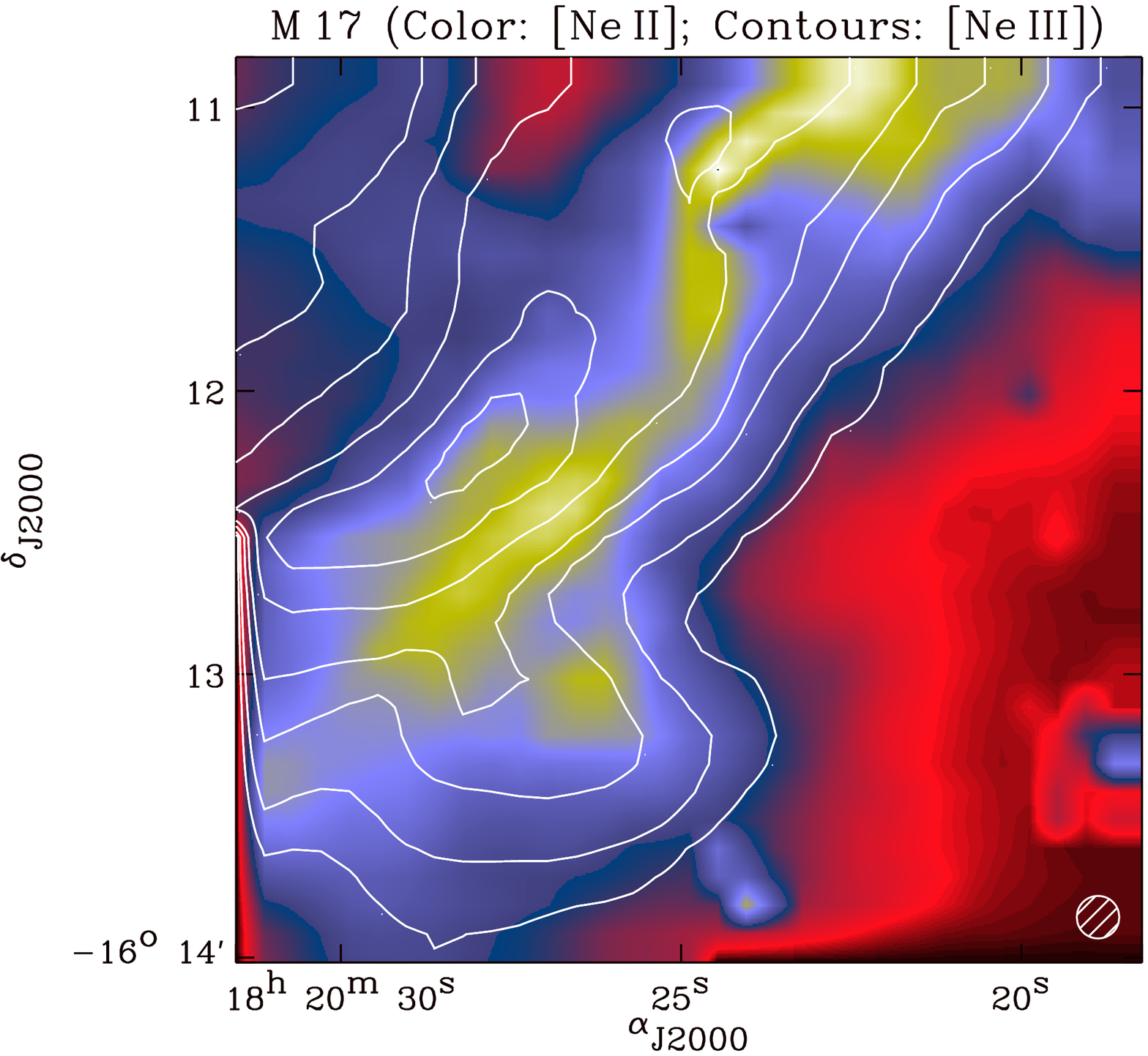} &
    \includegraphics[width=0.5\textwidth]{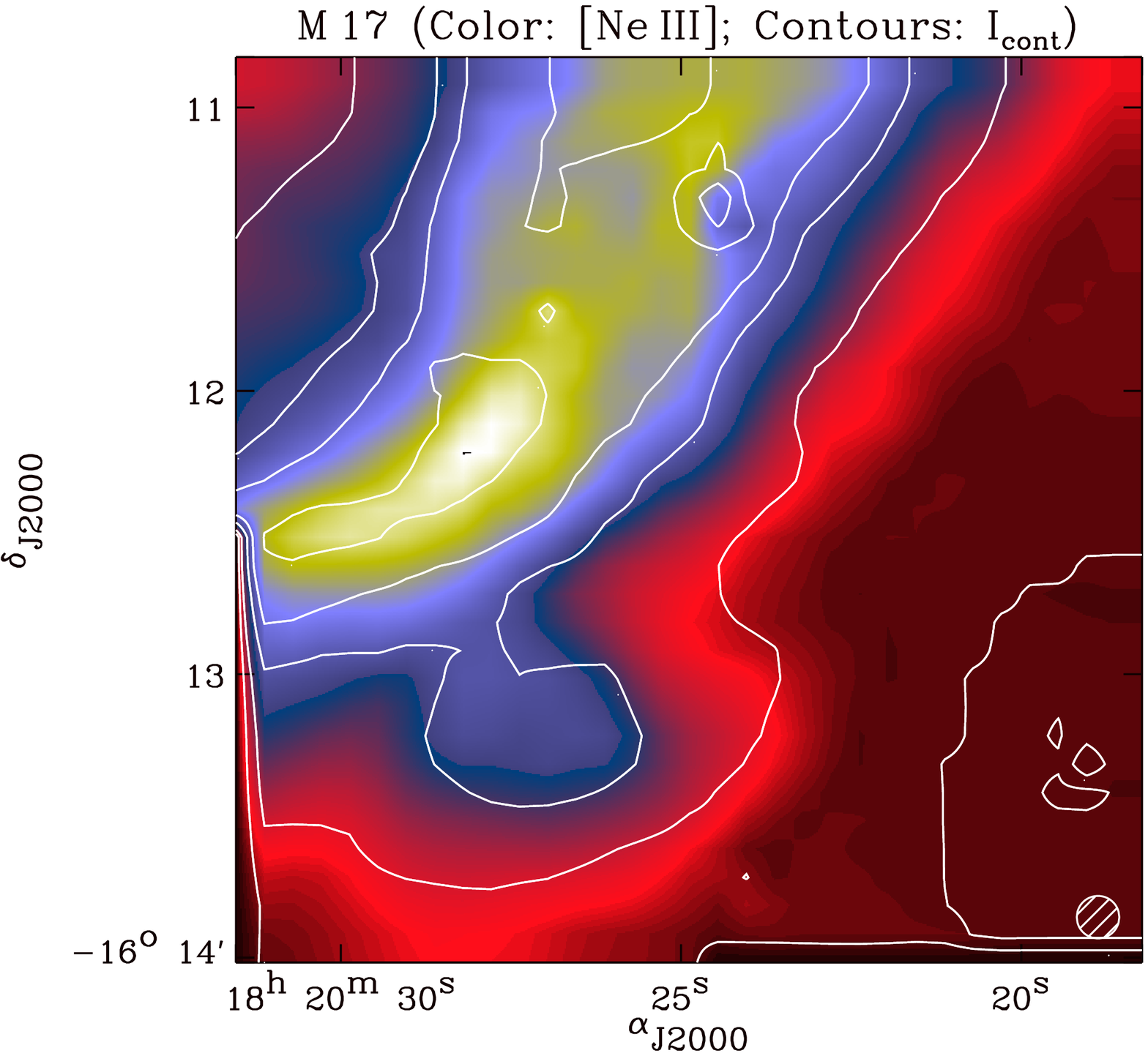} \\
    \includegraphics[width=0.5\textwidth]{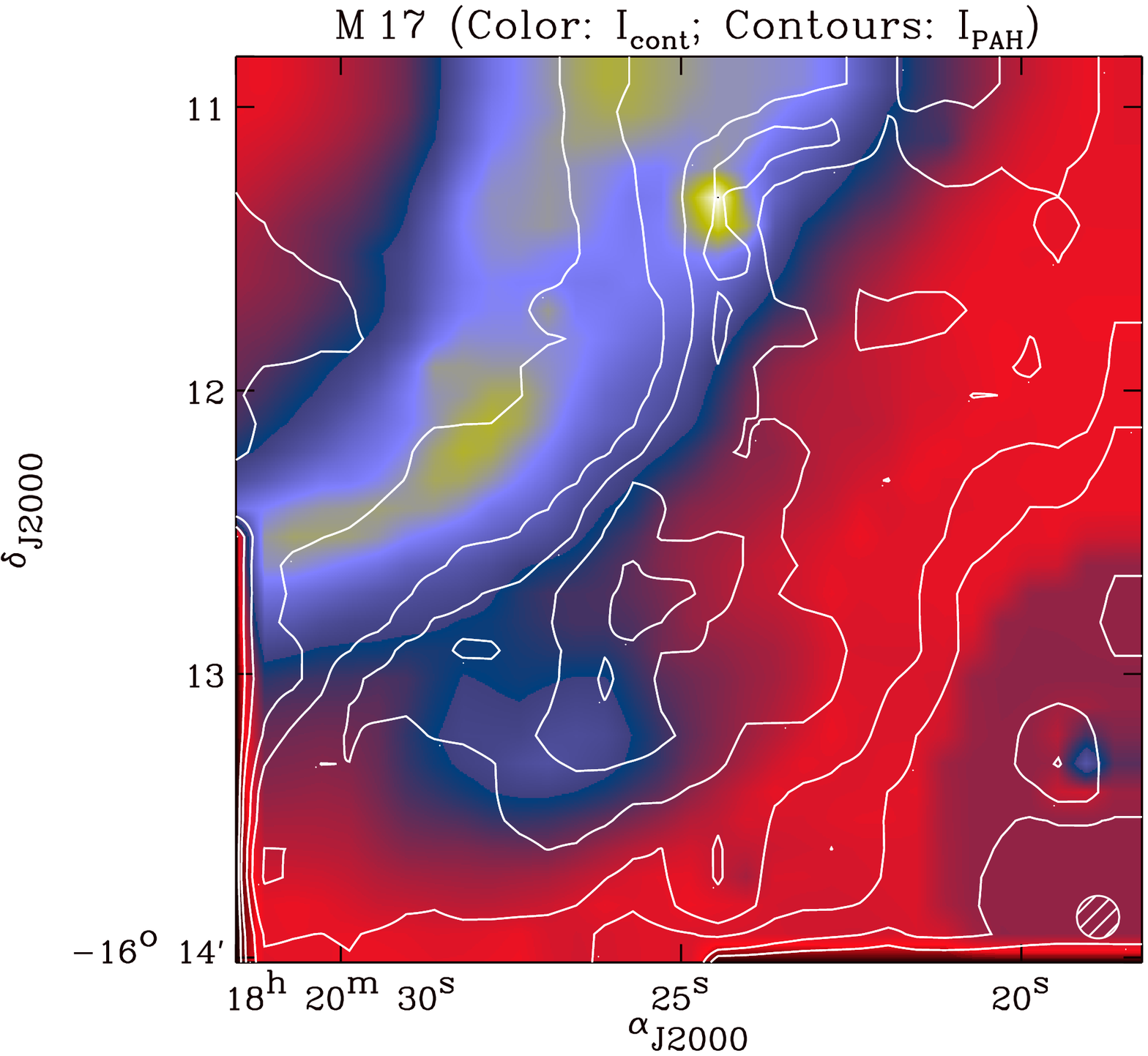} &
    \includegraphics[width=0.5\textwidth]{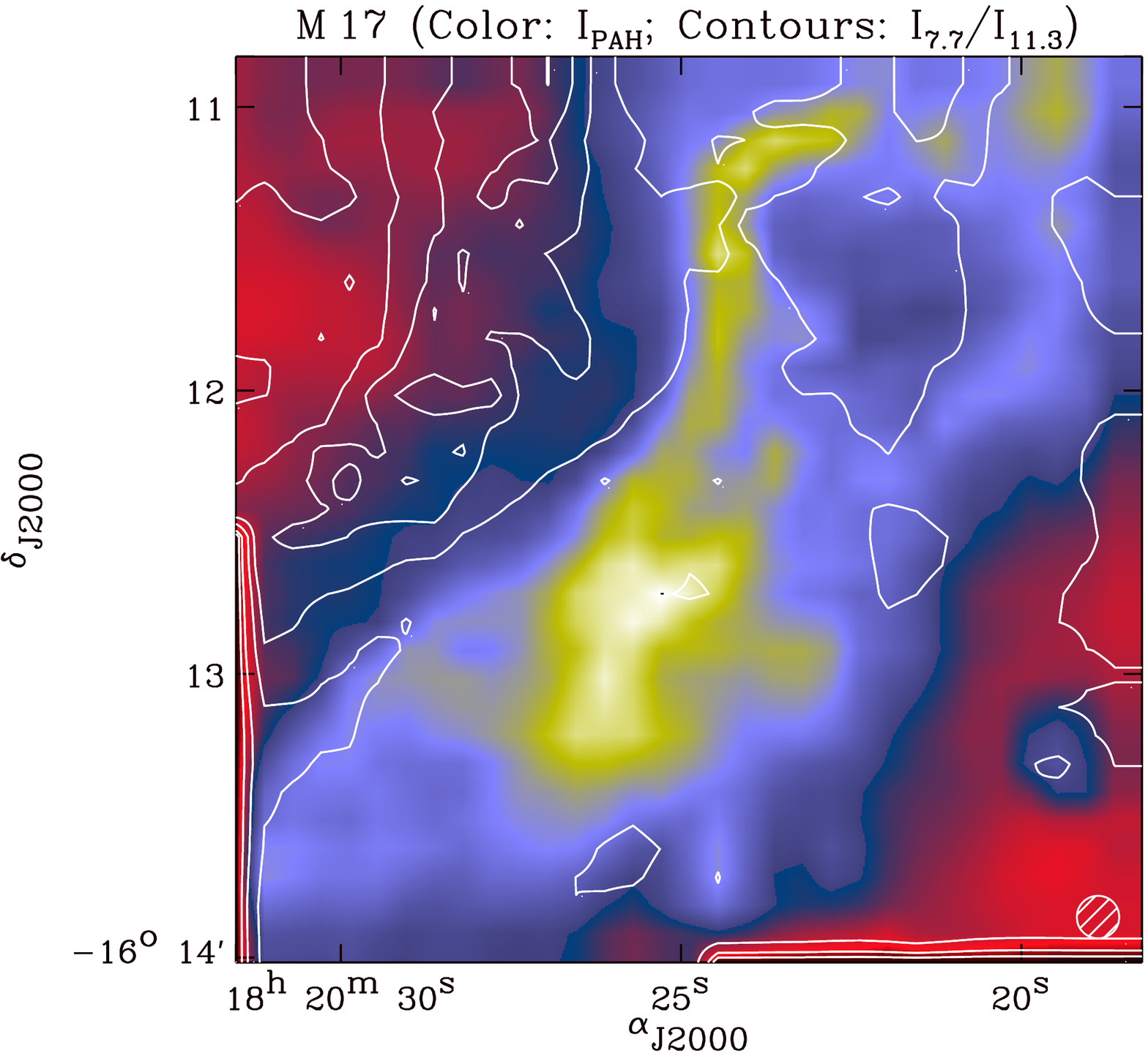} \\
  \end{tabular}}
  \caption{{\footnotesize\sl 
            Spatial distribution of the various mid-IR components within the
            Galactic PDR \M{17} \citep{cesarsky96,galliano08b}.
            Each image compares two spectral components; 
            the intensity increases from dark red to light yellow for the 
            color images; 
            the contour levels are 5, 10, 20, 30, 50, 70 and $90\%$ of the 
            peak intensity.
            The displayed components are the intensities of the {\rm\neiiline}
            and {\rm\neiiiline} lines, of the continuum ($\icont$, integrated 
            between 10 and 16$\mic$), of all the PAH bands $\iPah$, and the 
            $\ipah{7.7}/\ipah{11.3}$ ratio.
            The shaded circle in the bottom right corner of each figure is the
            beam size.
            }}
  \label{fig:m17}
\end{figure}

In general, high values of $\ipah{7.7}/\ipah{11.3}$ are associated with intense
star forming regions.
\reffig{fig:m17} shows images of several spectral components of the edge-on
Galactic photodissociation region (PDR) \M{17}.
On each image, the \hii\ region is located toward the top-left corner, and the
molecular cloud toward the bottom-right corner.
The edge-on view of this region allows us to see the different strata of 
species, as demonstrated by the top left panel of \reffig{fig:m17}.
This panel shows that the \neiii\ ions are distributed on a ridge located closer
to the \hii\ region than the \neii\ ions.
The top right panel shows that the \neiii\ and the very small grain intensity 
have similar spatial distributions, while the bottom left panel shows that the
PAHs are found more deeply inside the cloud, since they undergo 
photolysis/thermolysis close to the \hii\ region.
Finally, the bottom right panel shows that the $\ipah{7.7}/\ipah{11.3}$ ratio
peaks in the region where the \neiiiline\ and the continuum intensity $\icont$ 
are maximum.
Consequently, at the scale of an interstellar cloud, the 
$\ipah{7.7}/\ipah{11.3}$ ratio is maximum at the surface of the far-UV 
illuminated surface.

\begin{figure}[htbp]
  \centerline{
  \begin{tabular}{cc}
    \includegraphics[width=0.525\textwidth]{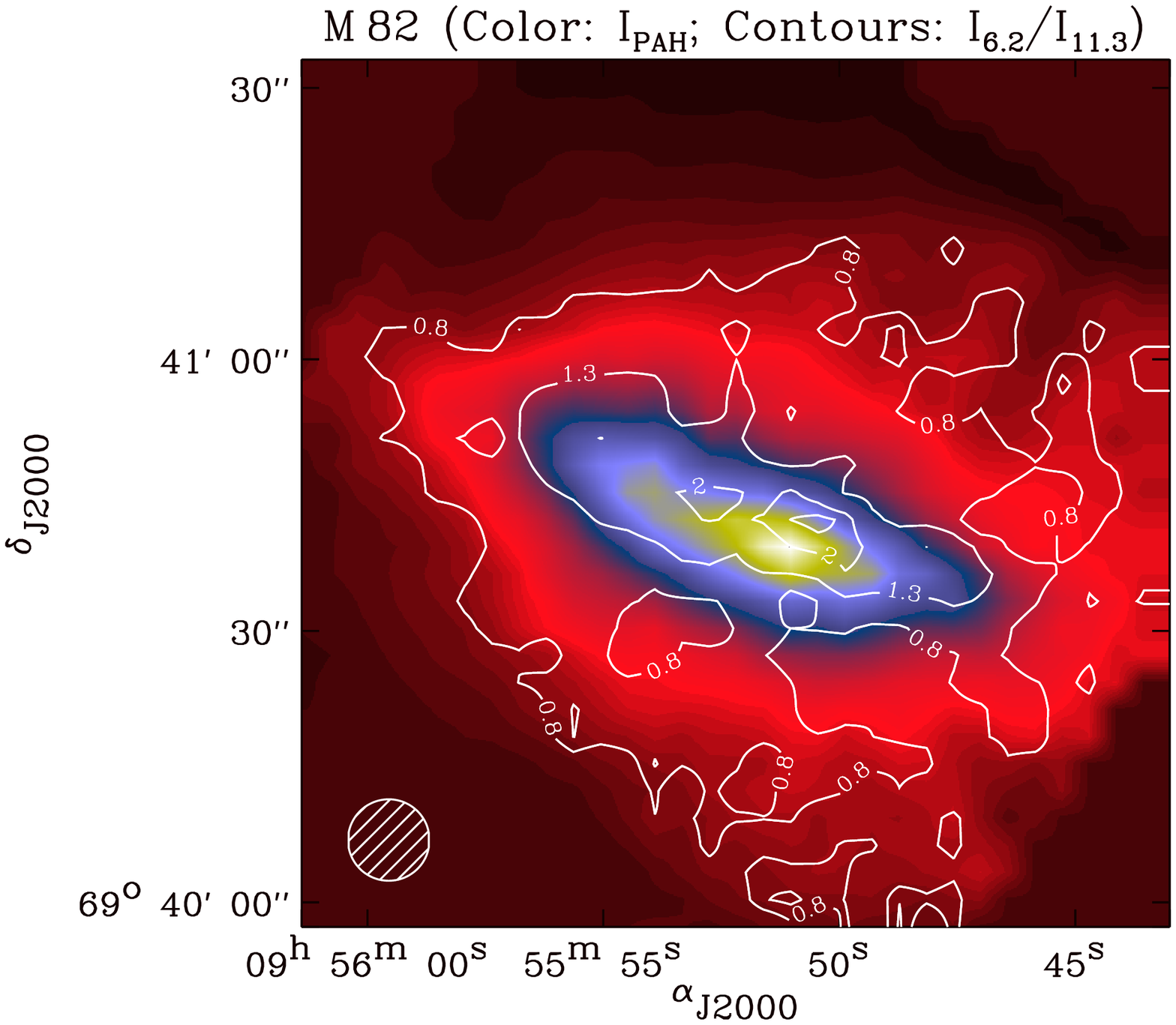} &
    \includegraphics[width=0.475\textwidth]{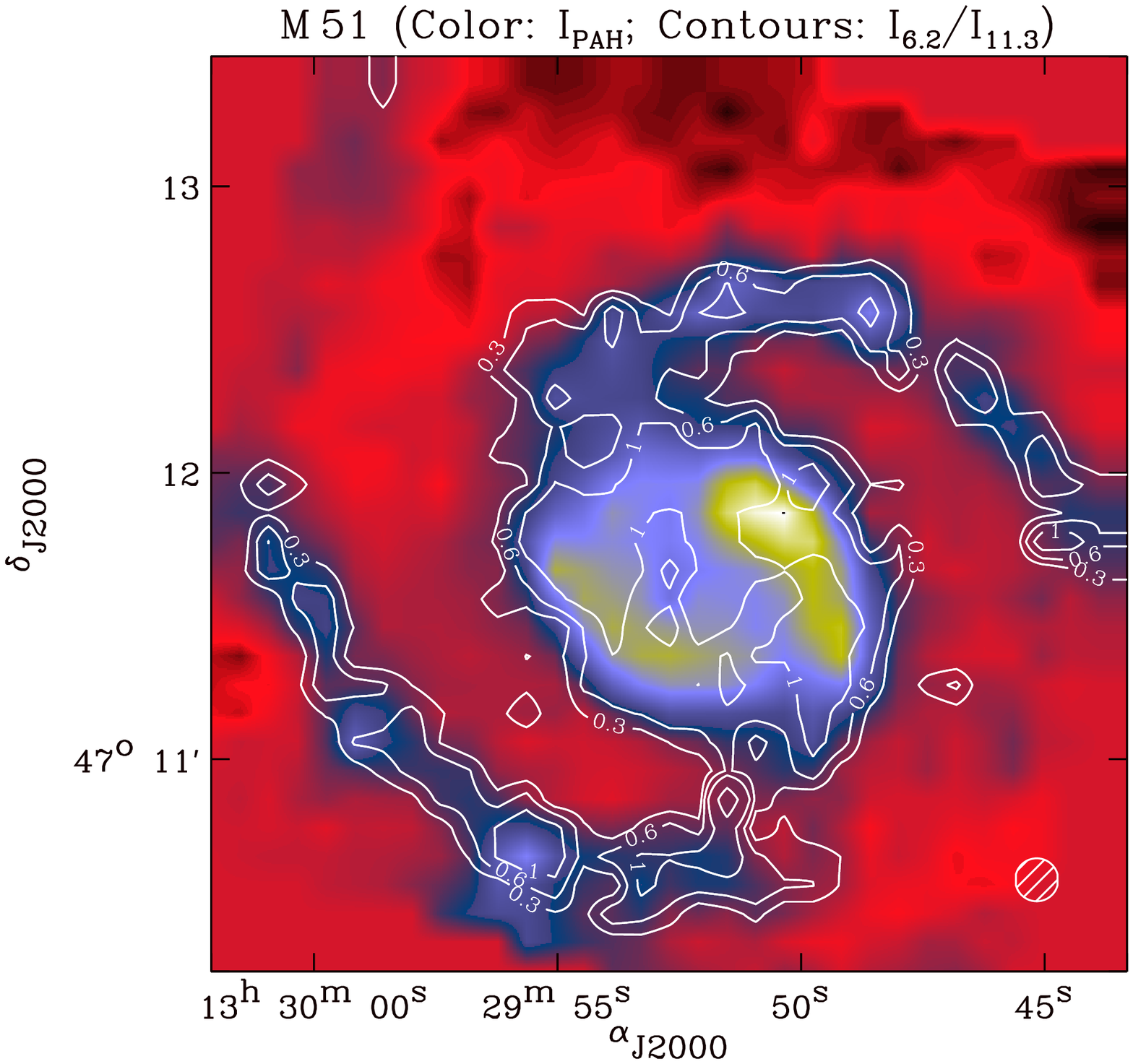} \\
  \end{tabular}}
  \caption{{\footnotesize\sl 
            Spatial distribution of the aromatic features in \M{82} (left panel)
            and \M{51} (right panel), from \citet{galliano08b}.
            The color images show the intensity of all PAH bands (increasing 
            from dark red to light yellow).
            The contours show the values of the $\ipah{6.2}/\ipah{11.3}$
            ratio.
            The shaded circle is the beam size.
            }}
  \label{fig:gal}
\end{figure}

\reffig{fig:gal} compares the spatial distributions of the intensity radiated 
by the aromatic bands to the $\ipah{6.2}/\ipah{11.3}$ ratio (following the same
trends than $\ipah{7.7}/\ipah{11.3}$), within two external galaxies, \M{82} and
\M{51}.
Contrary to \reffig{fig:m17}, the clouds are not resolved in this case.
These images indicate that the $\ipah{6.2}/\ipah{11.3}$ ratio is enhanced where
the PAH intensity is high, along the spiral arms and in the starburst regions.
Therefore, at the scale of a galaxy, the $\ipah{6.2}/\ipah{11.3}$ ratio is 
maximum in massive star forming regions.


\begin{figure}[htbp]
  \centering
  \includegraphics[width=\textwidth]{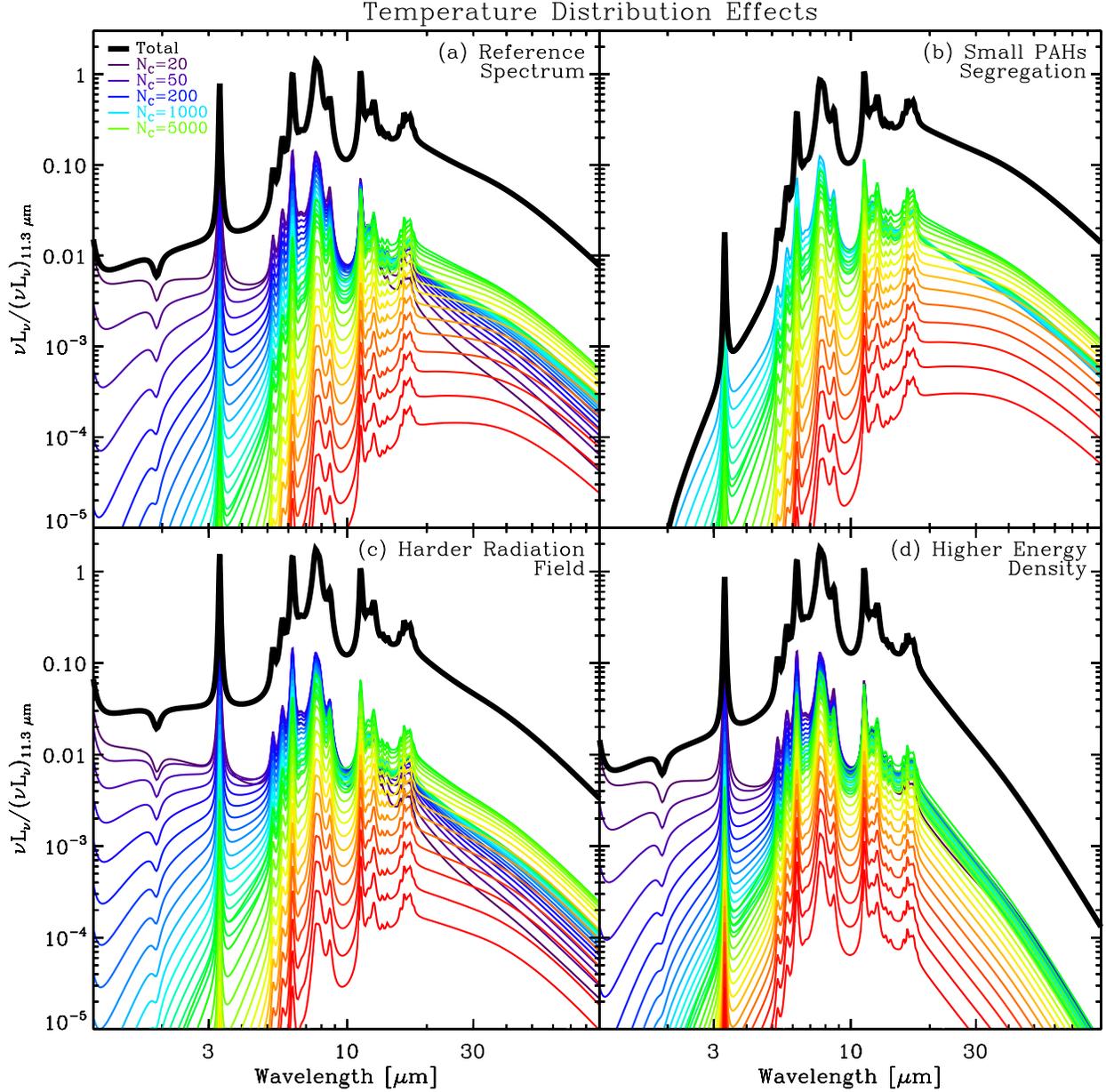}
  \caption{{\footnotesize\sl 
           Demonstration of the effects on the aromatic bands of changing the
           PAH temperature distribution.
           Each panel shows the theoretical spectrum of a collection of PAHs
           of various sizes $N_\sms{C}$ (number of carbon atoms; different 
           colors).
           The total spectrum integrated over the size distribution is the black
           thick line.
           \uline{Top-left panel:}
           we adopt the \citet{zubko04} size distribution, and the Galactic
           diffuse interstellar radiation field \citep[ISRF;][]{mathis83};
           this case serves as reference for the other panels.
           \uline{Top-right panel:}
           the size distribution is truncated; we exclude grains smaller than
           $N_\sms{C}=1000$.
           \uline{Bottom-left panel:} 
           the ISRF is harder (instantaneous burst of 0~Myr).
           \uline{Bottom-right panel:}
           the ISRF intensity is $\chi_\sms{ISRF}=10^5$ times higher.
           The absorption efficiencies are a mixture of 50$\%$ neutral and 
           50$\%$ ionized PAHs, from \citet{draine07}.
           }}
  \label{fig:PAHth1}
\end{figure}

\section{ORIGIN OF THE AROMATIC FEATURE VARIATIONS IN GALAXIES}

  \subsection{Inventory of the Physical Processes Potentially Affecting 
              the Band Ratios}
  \label{sec:PAHth}

PAHs are stochastically heated, hence their thermal emission spectrum 
is determined by their temperature distribution.
\reffig{fig:PAHth1} demonstrates the effects of changing this temperature 
distribution by varying different parameters.
\reffig{fig:PAHth1}a corresponds to the Galactic diffuse ISM.
\reffig{fig:PAHth1}b shows that the bands around 6.2, 7.7 and 8.6$\mic$
decrease relative to the 11.3$\mic$ band, if the small PAHs (smaller than 
$N_\sms{C}=1000$ carbon atoms) disappear.
Indeed, small PAHs fluctuate up to higher temperatures than larger ones.
Hence, their short wavelength features emit more intensely.
This case corresponds to environments where the small PAHs would photosublimate
preferentially.
\reffig{fig:PAHth1}c shows that by increasing the hardness of the interstellar
radiation field (ISRF) or equivalently by increasing the average stellar photon 
energy, PAHs fluctuate up to higher temperatures, therefore boosting the short 
wavelength features.
Finally, \reffig{fig:PAHth1}d shows that increasing the radiation density
decreases the time that a given PAH spend on average at cool temperatures.
However, this effect plays a role only for ISRFs which are 
$\chi_\sms{ISRF}\gtrsim 10^5$ times more intense than the Galactic one.

\begin{figure}[htbp]
  \centering
  \includegraphics[width=\textwidth]{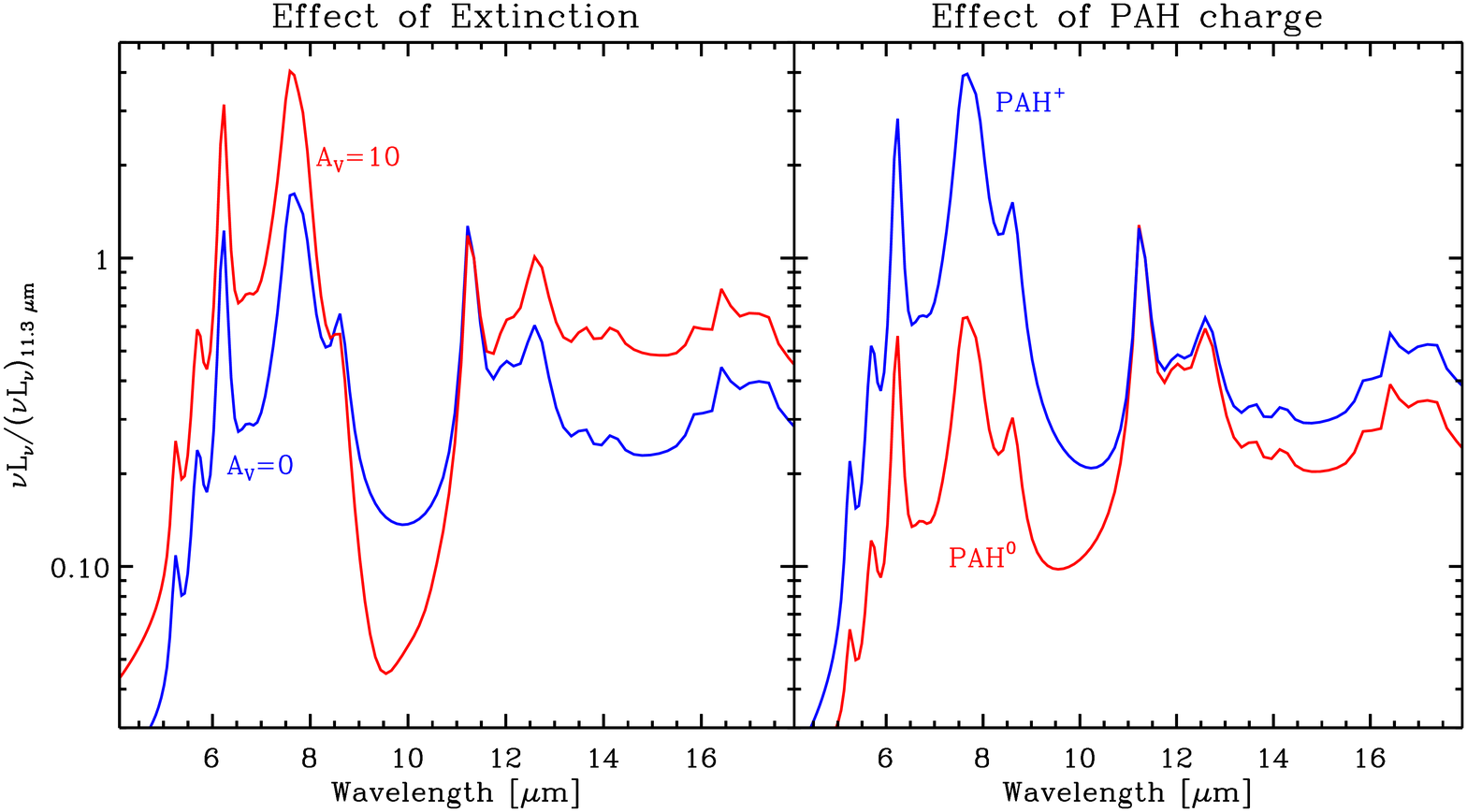}
  \caption{{\footnotesize\sl 
           Demonstration of the effects on the aromatic bands of the extinction
           and of the charge of the PAHs.
           Both panels show the theoretical spectra with the \citet{zubko04}
           size distribution, the \citet{draine07} absorption efficiencies,
           irradiated by the \citet{mathis83} ISRF.
           \uline{Left panel:}
           the reference spectrum (blue) is compared to the same spectrum
           extincted by a screen with an $A_\sms{V}=10$ (red).
           \uline{Right panel:}
           the red and blue spectra correspond to neutral and ionized PAHs.
           }}
  \label{fig:PAHth2}
\end{figure}

\begin{figure}[htbp]
  \centering
  \includegraphics[width=\textwidth]{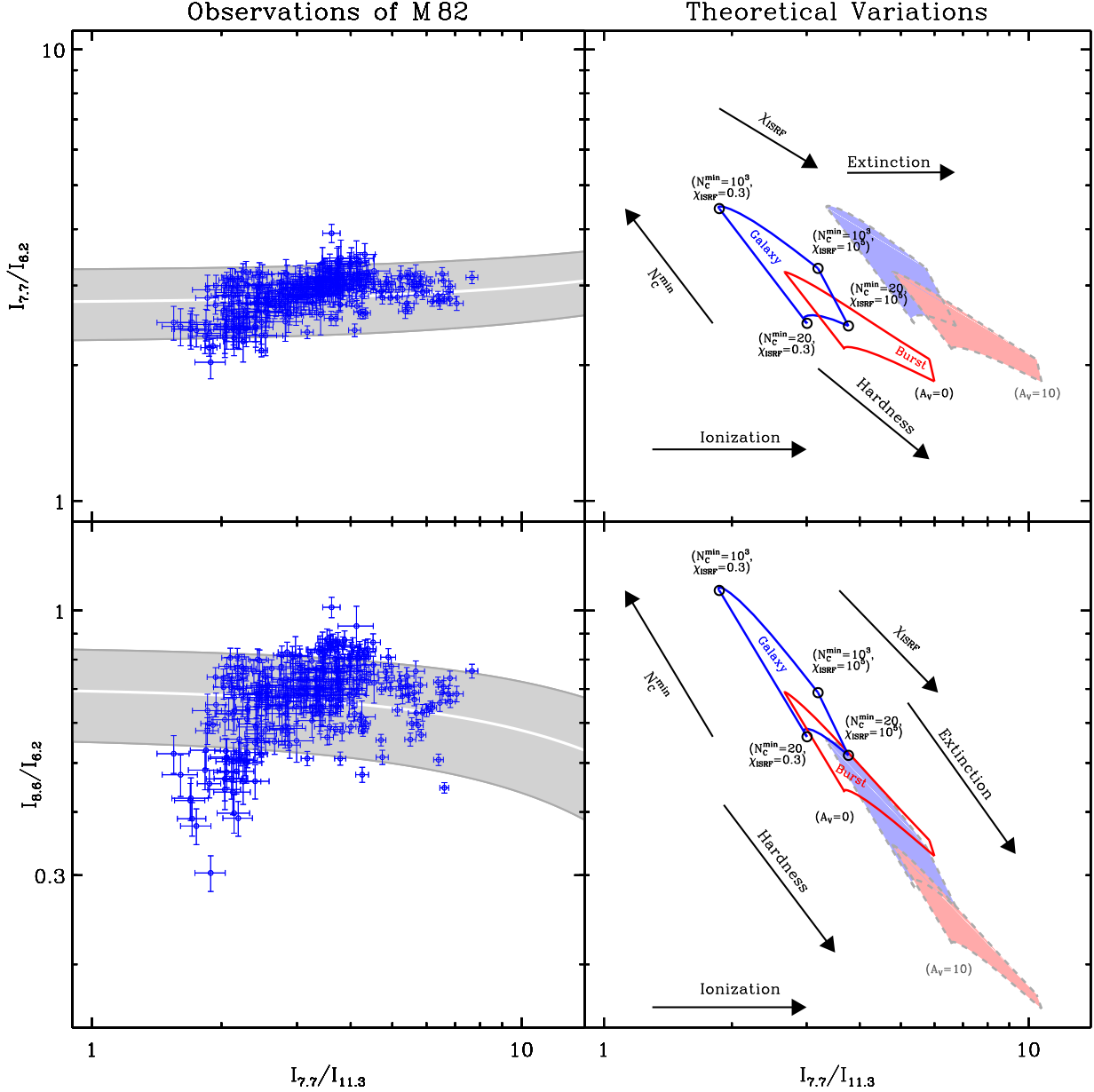}
  \caption{{\footnotesize\sl 
           Comparison between observed and theoretical trends of aromatic 
           feature intensity ratios.
           \uline{Left panels:}
           observed correlations within \M{82} (right panels of 
           \reffig{fig:correl});
           the grey stripes are line fits $\pm1\times\sigma$ to the 
           observations.
           \uline{Right panels:}
           presentation of the results of the modeling shown in 
           \reffigs{fig:PAHth1}-\ref{fig:PAHth2}.
           In each panel, the blue empty polygon with the circled corners
           shows the effect of varying both the minimum cut-off 
           PAH size, $N_\sms{C}^\sms{min}$, and the ISRF intensity,
           $\chi_\sms{ISRF}$, with the Galactic ISRF (labeled {\it Galaxy}).
           The values between parenthesis, close to each corners are the values
           of these two parameters at the corner.
           Second, the red empty polygon is the analog of the previous one,
           except that the ISRF is now a harder one (labeled {\it Burst}).
           Finally, the two filled polygons are the analogs of the two 
           previous ones, except that we applied a screen extinction with 
           $A_\sms{V}=10$.
           The various arrows show the sense of variation of the band ratios 
           with each parameter.
           }}
  \label{fig:comparison}
\end{figure}

\reffig{fig:PAHth2} demonstrates additional effects controling the aromatic 
feature ratios.
First, mid-IR extinction is dominated by a broad silicate feature centered 
around 9.7$\mic$.
This feature absorbs significantly the 8.6$\mic$ and 11.3$\mic$ bands,
but has little effect on the 6.2 and 7.7$\mic$ bands, as shown on the left panel
of \reffig{fig:PAHth2}.
Second, the 6 to 9$\mic$ features are enhanced for charged PAHs (right panel of 
\reffig{fig:PAHth2}), since they are attributed to C-C modes, while the 
11.3$\mic$ is attributed to peripheral C-H modes 
\citep[{e.g.}][]{allamandola89}.

  \subsection{The Dominant Role of the Fraction of Ionized PAHs}
  \label{sec:ionPAH}

\reffig{fig:comparison} compares the trend between band ratios observed within
\M{82} (similar trends are followed by the other sources; \reffig{fig:correl})
to theoretical trends.
The left panels of \reffig{fig:comparison} are another way to look at 
\reffig{fig:correl}.
Statistically, the observed $\ipah{7.7}/\ipah{6.2}$ and $\ipah{8.6}/\ipah{6.2}$
do not show significant variations and are uncorrelated with the 
$\ipah{7.7}/\ipah{11.3}$.
The right panels of \reffig{fig:comparison} represents the variations of the
aromatic features from the models presented in 
\reffigs{fig:PAHth1}-\ref{fig:PAHth2}.
What this figure shows is that the modification of the PAH temperature 
distribution (\refsec{sec:PAHth}) is not consistent with our observations,
since it would anticorrelate $\ipah{7.7}/\ipah{6.2}$ and $\ipah{8.6}/\ipah{6.2}$ 
with $\ipah{7.7}/\ipah{11.3}$.
On the other hand, the extinction produces the right trend between 
$\ipah{7.7}/\ipah{6.2}$ and $\ipah{7.7}/\ipah{11.3}$, but it anticorrelates
$\ipah{8.6}/\ipah{6.2}$ with $\ipah{7.7}/\ipah{11.3}$.
Finally, \reffig{fig:comparison} shows that the only effect which is consistent
with the observations is the variation of the fraction of ionized PAHs.

\begin{figure}[htbp]
  \centering
  \begin{tabular}{cc}
    \includegraphics[width=0.5\textwidth]{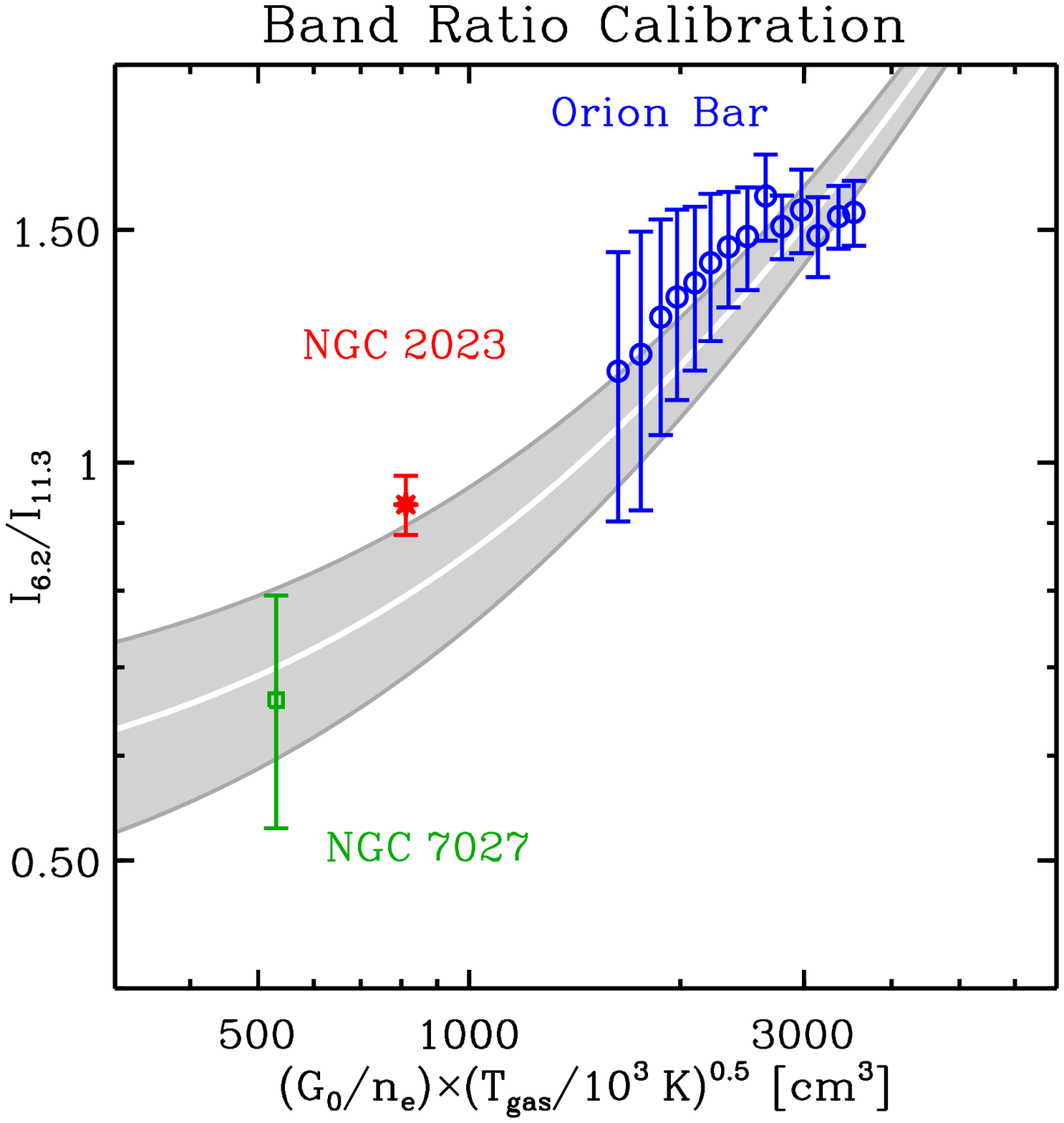} &
    \includegraphics[width=0.5\textwidth]{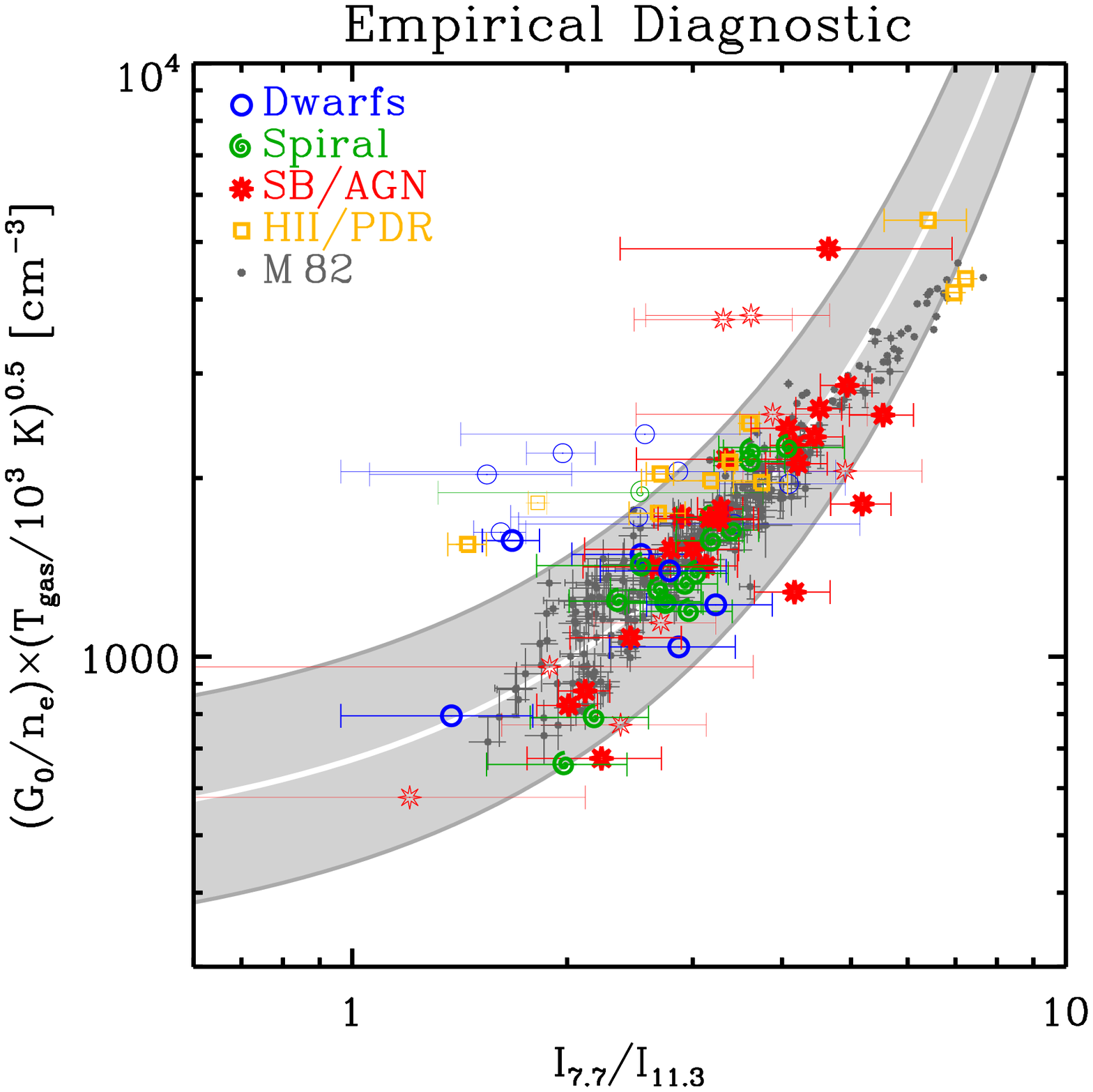} \\
  \end{tabular}
  \caption{{\footnotesize\sl 
            Empirical relation between the aromatic feature intensity ratio
            and the ionization-to-recombination ratio.
            \uline{Left panel:}
            calibration of the band ratio, using well calibrated sources;
            the values of $G_0$, $n_e$ and $T_\sms{gas}$ come from detailed
            PDR modeling.
            \uline{Right panel:}
            inversion of the right panel relation;
            the sources shown are the ones from \reffig{fig:correl}.
            The grey stripe is the line fit $\pm1\times\sigma$.
            }}
  \label{fig:g0ne}
\end{figure}

We emphasize that our sample contains mainly moderately obscured
star forming galaxies and Galactic regions, where AGNs are under-represented.
The effects that have been ruled out here may play a role for peculiar sources
like elliptical galaxies \citep{kaneda07}, LINERs \citep{smith07} or ULIRGs 
\citep{brandl06}.


\section{THE AROMATIC BANDS AS A DIAGNOSTIC TOOL OF THE PHYSICAL CONDITIONS}

  \subsection{Empirical Calibration of the Mid-Infrared Band Ratios}
  
The variations of aromatic feature intensity ratios are mainly controled by
the fraction of charged PAHs (\refsec{sec:ionPAH}).
This fraction of charged PAHs is directly related to the 
ionization-to-recombination ratio, $G_0/n_e\sqrt{T_\sms{gas}}$,
where $G_0$ is the UV field density, $n_e$ is the electron density, and
$T_\sms{gas}$ is the gas temperature.
Therefore the observed aromatic feature intensities can provide useful constraints
on the physical conditions in the region where they are emitting.

\begin{figure}[htbp]
  \centering
  \begin{tabular}{cc}
    \includegraphics[width=0.5\textwidth]{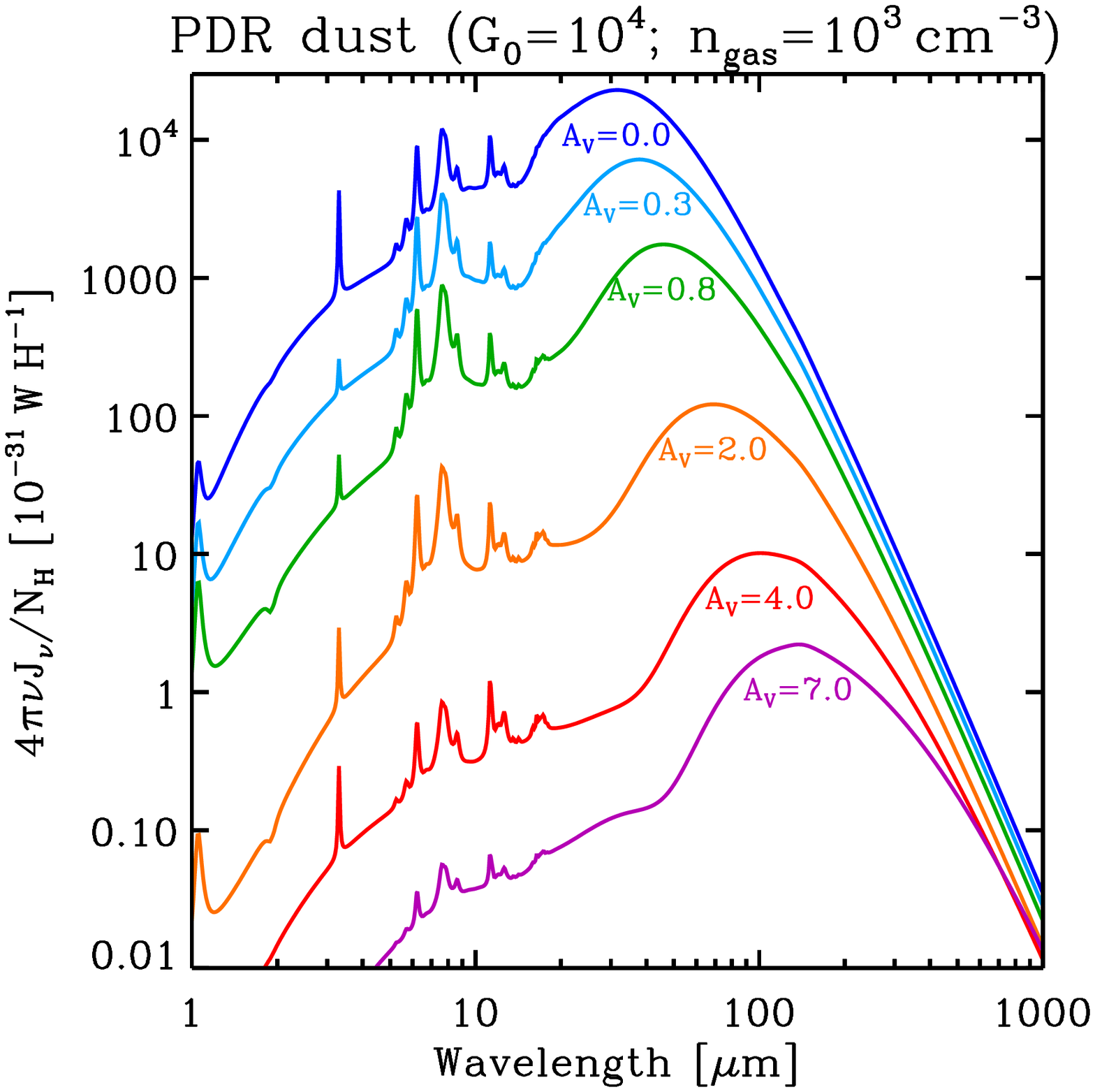} &
    \includegraphics[width=0.5\textwidth]{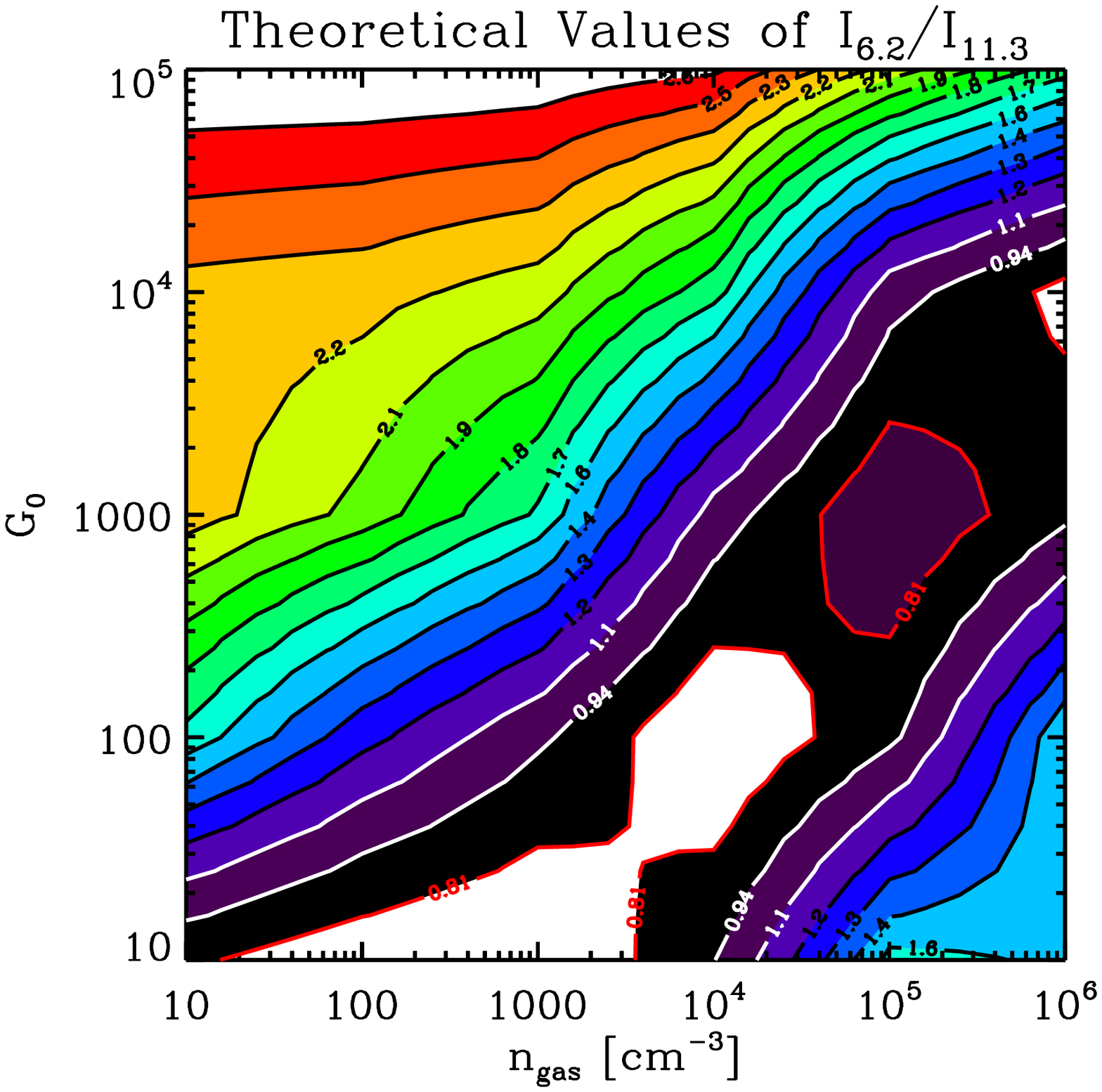} \\
  \end{tabular}
  \caption{{\footnotesize\sl 
            PAH and PDR self-consistent modeling, taking into account the
            radiative transfer, PAH stochastic heating and charge exchange between
            gas and dust \citep{galliano08d}.
            \uline{Left panel:}
            SEDs at different locations (corresponding to 
            different attenuations $A_\sms{V}$) within a given slab PDR having
            a density of $n_\sms{gas}=10^3\;\rm cm^{-3}$ and incident UV field
            of $G_0=10^4$ in Habing units.
            \uline{Right panel:}
            theoretical aromatic feature ratio diagram;
            each $(G_0,n_e)$ coordinate corresponds to a single face-on integrated
            PDR.
            }}
  \label{fig:PDR}
\end{figure}

We used the $G_0$, $n_e$ and $T_\sms{gas}$ values from the literature of a few 
well-studied sources, in order to provide an empirical calibration of the band 
ratio \citep{galliano08b}.
The left panel of \reffig{fig:g0ne} shows this calibration.
It demonstrates that the $\ipah{6.2}/\ipah{11.3}$ ratio (or the relative 
fraction of PAH$^+$) is high for either high $G_0$ (proportional to the 
ionization rate) or low $n_e/\sqrt{T_\sms{gas}}$ (proportional to the 
recombination rate).
This calibration can be applied to the sources in \reffig{fig:correl}
(right panel of \reffig{fig:g0ne}).

  \subsection{Detailed PDR Modeling of the Polycylic Aromatic Hydrocarbon 
              Emission}

The empirical relation of \reffig{fig:g0ne} can be generalised theoretically, 
using detailed PDR modeling.
The left panel of \reffig{fig:PDR} shows the dust spectral energy distributions
(SEDs) at different points within an homogeneous slab PDR.
The radiative transfer of the incident ISRF on the surface of the cloud is solved,
in combination with the photoionization, photodissociation and chemical reactions
within the cloud, using the PDR model of \citet{kaufman06}.
In particular, this PDR model takes into account the charge transfer between gas
and dust, and thus determines the fraction of charged PAHs at each point within 
the cloud.
Using the \citet{draine07} absorption efficiencies, we solved the stochastic 
heating of the PAHs and small grains for each $A_\sms{V}$ 
\citep[left panel of \reffig{fig:PDR};][]{galliano08d}.

The right panel of \reffig{fig:PDR} shows the value of the PAH band ratio 
integrated over each PDR of UV field $G_0$ and density $n_e$.
This type of diagram is usual for PDR gas lines.
The originality of this work is to extend it to dust features.
This figure is in qualitative agreement with \reffig{fig:g0ne}.
It shows that the $\ipah{6.2}/\ipah{11.3}$ ratio is constant over constant 
$G_0/n_e$ lines.
The $\ipah{6.2}/\ipah{11.3}$ is high at the top-left corner of the
panel, for high $G_0$ and low $n_e$, where the PAHs are mainly positively charged.
The $\ipah{6.2}/\ipah{11.3}$ is also high at the botton-right corner of the
panel, for low $G_0$ and high $n_e$, since the PAHs are mainly negatively charged.
\seppar

The present study has developed the observed mid-IR features
as a quantitative tool to probe the physical conditions 
({e.g.} $G_0/n_e\times\sqrt{T_\sms{gas}}$) in the emitting regions.
We expect that this study will be of fundamental value for the interpretation 
of \spitz\ data as well as future \sofia\ and \jwst\ observations of galaxies in 
the nearby and early universe.


\bibliographystyle{/Users/Fred/Documents/Astro/TeXstyle/Packages_AAS/aas}
\bibliography{/Users/Fred/Documents/Astro/TeXstyle/references}

\end{document}